\documentclass[twocolumn,showpacs,pra,aps]{revtex4-1}

\usepackage[T1]{fontenc}     %Output what you want e.g., é, ł, a, ü
\usepackage[british]{babel}  %Do hyphenation according to british english
\usepackage{lmodern}  %Joe's font

\usepackage[usenames,dvipsnames]{color} %More color choices

\usepackage{graphicx} % Package to insert exteral figures
\usepackage{amsmath,amssymb,amsthm,bm,amsfonts,mathrsfs,bbm} %Usefull math packages
\usepackage{xspace}  %Useful to add space in macros
\usepackage{pgfplots}  %Useful to plot graphs with latex
\usepackage[colorlinks,linkcolor=magenta,urlcolor=blue]{hyperref}  %Hyperlinks (pink, green, blue)
\usepackage{mathtools}

\DeclareMathOperator{\tr}{tr}

\newcommand{\bra}[1]{\left\langle #1 \right|}
\newcommand{\ket}[1]{\left| #1 \right\rangle}

\newcommand{\ba}{\begin{eqnarray}}
\newcommand{\ea}{\end{eqnarray}}
\newcommand{\ban}{\begin{eqnarray*}}
\newcommand{\ean}{\end{eqnarray*}}

\usepackage{ifxetex,ifluatex}
\ifnum 0\ifxetex 1\fi\ifluatex 1\fi=0 % if pdftex
  \usepackage[T1]{fontenc}
  \usepackage[utf8]{inputenc}
\else % if luatex or xelatex
  \ifxetex
    \usepackage{mathspec}
  \else
    \usepackage{fontspec}
  \fi
  \defaultfontfeatures{Ligatures=TeX,Scale=MatchLowercase}
\fi
\IfFileExists{upquote.sty}{\usepackage{upquote}}{}
\IfFileExists{microtype.sty}{%
\usepackage[]{microtype}
\UseMicrotypeSet[protrusion]{basicmath} % disable protrusion for tt fonts
}{}
\PassOptionsToPackage{hyphens}{url} % url is loaded by hyperref
\usepackage{longtable,booktabs}
\IfFileExists{footnote.sty}{\usepackage{footnote}\makesavenoteenv{long
table}}{}
\IfFileExists{parskip.sty}{%
\usepackage{parskip}
}{% else
\setlength{\parindent}{0pt} \setlength{\parskip}{6pt plus 2pt
minus 1pt} }
\providecommand{\tightlist}{%
  \setlength{\itemsep}{0pt}\setlength{\parskip}{0pt}}
\ifx\paragraph\undefined\else
\let\oldparagraph\paragraph
\renewcommand{\paragraph}[1]{\oldparagraph{#1}\mbox{}}
\fi \ifx\subparagraph\undefined\else
\let\oldsubparagraph\subparagraph
\renewcommand{\subparagraph}[1]{\oldsubparagraph{#1}\mbox{}}
\fi

\makeatletter
\def\fps@figure{htbp}
\makeatother

\newcommand{\comm}[1]{\quad\quad\text{--- #1}}
\newcommand{\Z}{\mathbb{Z}}
\newcommand{\mat}[2]{\mathcal{M}_{#1\times #2}}
\newcommand{\opt}{L}
\newcommand{\digits}[1]{\overline{#1}}

\begin{document}

%\title{$K_G(3)<K_G(4)$ and qubit witnesses}
%\title{$K_G(3)<K_G(4)$}

% journal title:
\title{Qutrit witness from the Grothendieck constant of order four}

\author{P\'eter Divi\'anszky}
\affiliation{Institute for Nuclear Research, Hungarian Academy of
Sciences, H-4001 Debrecen, P.O. Box 51, Hungary}
\author{Erika Bene}
\affiliation{Institute for Nuclear Research, Hungarian Academy of
Sciences, H-4001 Debrecen, P.O. Box 51, Hungary}
\author{Tam\'as V\'ertesi}
\affiliation{Institute for Nuclear Research, Hungarian Academy of
Sciences, H-4001 Debrecen, P.O. Box 51, Hungary}

\date{\today}  %Date today

\begin{abstract}
In this paper, we prove that $K_G(3)<K_G(4)$, where $K_G(d)$
denotes the Grothendieck constant of order $d$. To this end, we
use a branch-and-bound algorithm commonly used in the solution of
NP-hard problems. It has recently been proven that $K_G(3)\le
1.4644$. Here we prove that $K_G(4)\ge 1.4841$, which has
implications for device-independent witnessing dimensions greater than two.
Furthermore, the algorithm with some modifications may find
applications in various black-box quantum information tasks with
large number of inputs and outputs.
\end{abstract}

\maketitle

\section{Introduction}
\label{sIntro}

The Grothendieck constant $K_G$ \cite{grot} is an enigmatic
constant arising in Banach space theory~\cite{pisier} with several
recent applications in communication complexity~\cite{commc} and
algorithms~\cite{alg1,alg2}. It is known to be in the range
$1.6769<K_G<1.7823$, however, its exact value is still unknown.
The lower bound above has been given by Davie and
Reeds~\cite{DavieReeds}, and the upper bound is due to
Krivine~\cite{Krivine}. There is a refined version of the
Grothendieck constant, the Grothendieck constant of order $d$,
denoted by $K_G(d)$. The definition of $K_G(d)$ for any finite
$d\ge 2$ is given below. Note that the original constant $K_G$ is
recovered for $d\rightarrow\infty$.

Let us first define $L(M)$ by the optimization problem
\begin{equation}
\label{LM}
L(M)=\max_{a_i=\pm1,b_j=\pm1}\sum_{i=1}^n\sum_{j=1}^n M_{ij}a_ib_j
\end{equation}
over all possible signs of $a_i,b_j$, $i,j=1,\ldots,n$, where $M=(M_{ij})$ is an arbitrary $n\times n$ real-valued matrix. The optimization problem~(\ref{LM}) is called $K_{n,n}$-quadratic programming in the computer science literature~\cite{raga}. This is known to be an NP-hard problem in the parameter $n$~\cite{Pitowsky08}. The Grothendieck inequality~\cite{grot,finch} states that
\begin{equation}\label{grot1}
\frac{\sum_{i=1}^n\sum_{j=1}^n M_{ij}\vec a_i\cdot\vec b_j}{L(M)}\le K(d),
\end{equation}
for all unit vectors $\vec a_1, \vec a_2,\ldots,\vec a_n\in\mathbb{R}^d$ and $\vec b_1, \vec b_2,\ldots,\vec b_n\in\mathbb{R}^d$, where $K(d)$ is a universal constant for a fixed $d$. The smallest value of this constant $K(d)$ such that the inequality still holds is called the Grothendieck constant of order $d$, which we denote by $K_G(d)$. Recall that $K_G=\lim_{d\rightarrow\infty}K_G(d)$.

Despite efforts, the value of the constant $K_G(d)$ is not known
in general and its exact value is only known for $d=2$:
$K_G(2)=\sqrt 2$ \cite{Krivine,CHSH}. For larger $d$, there
appeared better-and-better lower
bounds~\cite{DavieReeds,FR94,vertesi,hua1,hua2,modgilbert} and
upper bounds~\cite{Krivine,bra11,KG3upper} to $K_G(d)$ in the
literature.

Our goal is to improve on existing lower bounds for $K_G(d)$ in
the case of small dimensions $d$. Note that according to the
definition~(\ref{grot1}), a lower bound to $K_G(d)$ arises by
giving an explicit matrix $M$ and explicit unit vectors $\vec a_i$
and $\vec b_j$ in dimension $d$. Denoting by
\begin{equation}
\label{QMd}
Q(M,d)=\sum_{i=1}^n\sum_{j=1}^n M_{ij}\vec a_i\cdot\vec b_j
\end{equation}
the nominator in the left-hand side of the inequality~(\ref{grot1}), we get the following lower bound:
\begin{equation}
\label{KGlower}
\frac{Q(M,d)}{L(M)}\le K_G(d).
\end{equation}

Currently, the best-known lower bound is given by $K_G(4)\ge
1.4456$ in Ref.~\cite{hua1}. In this paper, we improve on this bound up to $K_G(4)\ge
1.4841$. Since $K_G(3)$ is known to be smaller than $1.4644$~\cite{KG3upper}, the
strict relation $K_G(3)<K_G(4)$ follows. As a by-product, we also
improve the best lower bound on $K_G(3)$. To this end, we combine
the so-called distance algorithm~\cite{gilbert,modgilbert} with a
branch-and-bound algorithm~\cite{BB}. Let us note that there is a
connection between the Grothendieck constant of order $d$ and the
nonlocality of XOR games~\cite{xor}. This link has been
established by Tsirelson~\cite{Tsirelson87,Tsirelson92} and
further expanded in Ref.~\cite{acin06}. Our result $K_G(3)<K_G(4)$
will have implications in this direction as well entailing a so-called
dimension witness for systems beyond qubits~\cite{dimwit}. Since both the distance and the branch-and-bound methods
have been applied independently in versatile schemes, we believe
that together they will find applications in other nonlocality
scenarios and large-scale quantum information tasks as well.

The paper is organized as follows. In Sec.~\ref{sBB}, we introduce
the branch-and-bound (B\&B) algorithm to solve problem~(\ref{LM})
and we also present test cases showing its performance for large
$n$ matrix dimensions. In Sec.~\ref{sImp}, the lower bounds are
improved both for $K_G(3)$ and $K_G(4)$. In particular, a $92\times92$ matrix $M$ is constructed in
Sec.~\ref{ssKG4dir} showing that $K_G(4)\ge 1.4731$, which is
further improved to $K_G(4)\ge 1.4841$ in Sec.~\ref{ssKG4dist} by
invoking the distance algorithm. Similarly, it is shown in
Sec.~\ref{ssKG3dist} using a $90\times 90$ matrix that $K_G(3)\ge
1.4359$. Note that the best lower bound so far was $K_G(3)\ge
1.4261$, presented in Ref.~\cite{modgilbert}. The connection with
nonlocal quantum correlations and the implications for
device-independent dimensions witnesses are discussed in
Sec.~\ref{sBell}. The paper ends with conclusions in
Sec.~\ref{sDisc}.

\section{The Branch-and-Bound (B\&B) algorithm}
\label{sBB}

\subsection{Description of the algorithm}
\label{ssDes}

Let us recall from the introduction that a good lower bound to
$K_G(d)$ requires a suitable $n\times n$ matrix $M$ along with a specific
arrangement of unit vectors $\vec a_i,\vec b_j$. Armed with these,
we also need a method which is able to efficiently evaluate the
maximum $L(M)$ in formula~(\ref{LM}) for large matrix dimensions
$n$. In this section, we propose a solution based on a
branch-and-bound technique~\cite{BB}, which is feasible on a
standard computer up to $n\sim90$.

It is known that assuming the Unique Games Conjecture~\cite{khot},
it is NP-hard to approximate the above problem to any factor
better than the Grothendieck constant $K_G$ \cite{raga}. Actually,
if $M$ is the Laplacian matrix of a graph then the maximum in
(\ref{LM}) coincides with the value of the maximum cut of this
graph. The maximum cut problem is one of $21$ NP-complete problems
of Karp~\cite{karp}.

Notice that the optimization problem~(\ref{LM}) reduces to the
following problem (where the $n\times n$ matrix $M$ is the input
to the problem):
\begin{equation}
\label{LM2}
L(M)=\max_{a_i=\pm1}\sum_{j=1}^n\left|\sum_{i=1}^n M_{ij}a_i\right|,
\end{equation}
where maximization is performed over all possible $\pm 1$ signs of $a_i$, $i=1,\ldots,n$. This
reformulation of the problem allows us to eliminate variables
$b_j$ from the optimization. Note, however, that a brute-force
search evaluation of this problem becomes infeasible already for
relatively small $n$, as one has to compute all the $2^{n-1}$ distinct cases.
Such a brute force technique was used in
Refs.~\cite{modgilbert,montina}, and the biggest $n$ one could
afford (in a reasonable time) on a normal desktop PC was $n=42$.

In contrast, the B\&B algorithm is able to cope with generic matrices $M$ with dimensions up to $n\sim 90$ in a reasonable time as it will be discussed next. In our specific problem~(\ref{LM2}), the B\&B
algorithm performs a systematic enumeration of candidate solutions
for $a_i$, $i=(1,\ldots,n)$ by means of state space
search~\cite{BB}: We can think of our set of candidate solutions
as a rooted binary tree with the full set at the root. Let us
label a given branch by a particular choice of $\pm1$ signs of
$\{a_1,a_2,\ldots,a_n\}$ variables. Then the algorithm explores
branches of this tree representing subsets of the solution set.
Before enumerating the candidate solutions of a branch, the branch
is checked against estimations of upper bounds on the optimal
solution, and the branch is removed if it cannot produce a better
solution than the best one found so far by the algorithm.

The estimation of the upper bound is based on the following
inequality. Let us fix values of $a_i=\{+1,-1\}, i=1,\ldots,k$,
corresponding to the level $k$ of the branching tree. Then we have
the following upper bound:
\begin{align} \label{grotupp}
&\max_{a_{k+1},\ldots,a_n}\sum_{j=1}^n\left|\sum_{i=1}^n M_{ij}a_i\right| \nonumber\\
&\le \sum_{j=1}^n\left|\sum_{i=1}^k M_{ij}a_i\right|+\max_{a_{k+1},\ldots,a_n}\sum_{j=1}^n\left|\sum_{i=k+1}^n M_{ij}a_i\right|,
\end{align}
where maximization is carried out over all possible $\pm 1$ signs
of $a_{k+1},\ldots,a_n$. It is noted that the first term on the
right-hand side of Eq.~(\ref{grotupp}) has some fixed value, which
for consecutive $k$'s can be computed at low cost by reusing
results from previous computations.

An efficient upperbounding is a crucial part of
the algorithm, since without discarding branches, the technique
traces back to a brute force search of all possible solutions,
which amounts to evaluating $2^{n-1}$ solutions growing
exponentially with $n$. According to the upper
bound~(\ref{grotupp}), the decision about which branches to remove
can be taken quickly. We refer the interested reader to the
Appendix~\ref{appa} for a detailed technical description of the
algorithm along with simple illustrative examples.

Let us stress that the B\&B algorithm described above allows us to
give an exact value for the problem~(\ref{LM}) if all entries of
matrix $M$ are integers. This algorithm was implemented in Haskell
language and is available in the webpage~\cite{github}. The code
performs exact integer arithmetic and includes assembly code in
certain crucial parts to boost the computation.

\subsection{Numerical tests}
\label{ssNum}

\begin{figure}[htb!]
%\centering
%\includegraphics[width=1.3\columnwidth]{fig1.png}
% [trim={<left> <lower> <right> <upper>}]
\includegraphics[width=1.0\columnwidth,trim={3.0cm 9.0cm 3.5cm 10.0cm},clip]{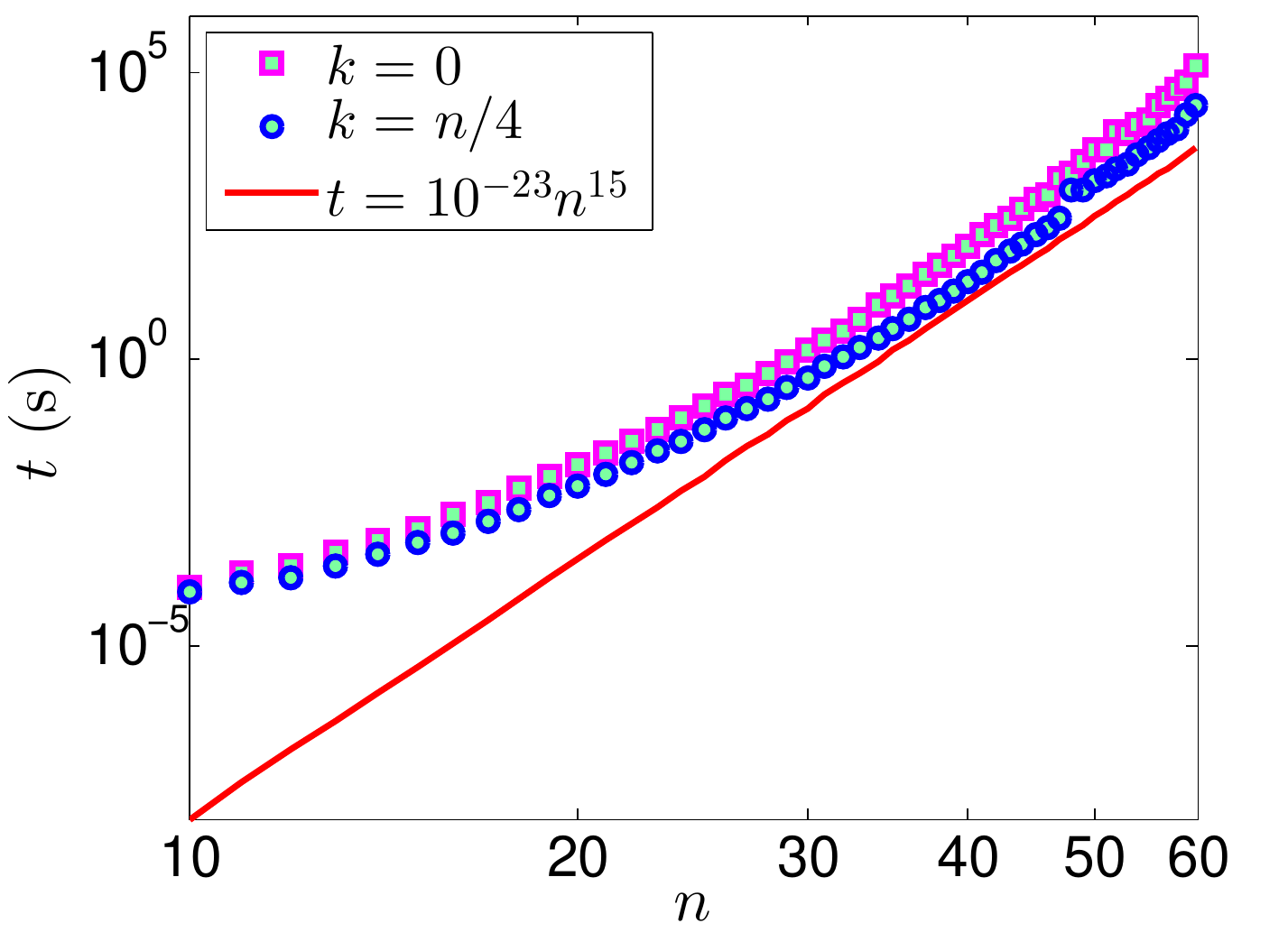}
\caption{Time required to compute $L(M)$ as a function of the matrix dimension $n$ of $M$. The plot has a log-log scale. The purple square and blue circle markers correspond to the curves with parameter $k=0$ and $k=n/4$, respectively. The red line is for the sake of comparison.}
\label{fig1}
\end{figure}

In this section some benchmark tests are presented. We generated
$n\times n$ random $M$ matrices for a given $n$, where the
integer coefficients of the matrix $M$ were chosen within the range
$[-100,+100]$ uniformly at random. After averaging over 1000
random matrices for a fixed $n$, we plot the time required for
computing $L(M)$ using the B\&B algorithm as a function of $n$ in
the range $10\le n\le 60$ (it is noted that for $50<n\le60$,
the average was taken over only 30 matrices to save computation time).
The code was run on a single core of a standard desktop PC.
Fig.~\ref{fig1} shows the performance of the B\&B algorithm on a
log-log plot. Note that there is a parameter $k$ in the algorithm,
which designates the level of the tree above which all nodes are
forced to be visited. In this way we can save computation time, since less decisions
have to be taken about discarding branches from the tree.
To our experience, choosing $k\sim n/4$ gives the best
performance. In Fig.~\ref{fig1}, we plotted both cases $k=0$ and $k=n/4$,
demonstrating that $k=n/4$ is indeed superior to $k=0$.

For the sake of comparison, we also plotted the line
$t=10^{-23}n^{15}$ (shown in red). As one can observe, the
performance of the B\&B algorithm can be well approximated with a
power-law behavior in the range displayed ($n\le60$). By
extrapolating the curve $t=10^{-23}n^{15}$ up to $n=90$, we get a
running time in the range of month (carried out on a single core).
However, for higher $n$ and generic matrices $M$, we expect an
exponentially growing behavior due to the NP-complete feature of
the problem. Indeed, one can easily construct specific matrices $M$ for which there is no saving in the running time compared to the brute-force technique which has exponential scaling with $n$. Such a matrix maybe built up of $(n/2)$ Clauser-Horne-Shimony-Holt expressions~\cite{CHSH} distributed between $n$ settings of Alice and Bob (where $n$ is even). In this case, the number of different $a_i=\pm 1$, ($i=1,2,\ldots,n$) strategies attaining $L(M)$ in Eq.~(\ref{LM2}) grows exponentially with $n$. In such a case the number of discarded branches is limited and the performance of the algorithm eventually goes back to that of a brute force search.

On the other hand, let us also stress that the memory complexity of the
algorithm is low. It equals the size of the input matrix $M$ of
the problem, which is $O(n^2)$.

%This is also apparent due to the $k=n/4$ curve which has to scale in $n$ at least as $O(\exp{(n/4)})$.

\section{Improving the lower bound on $K_G(4)$ and $K_G(3)$}
\label{sImp}

Our task is to come up with a good lower bound to $K_G(d)$, which according to (\ref{KGlower}) amounts to finding a suitable arrangement of unit vectors $\vec a_i,\vec b_j$ in $\mathbb{R}^d$ for $i=1,\ldots,n$ and an $n\times n$ dimensional matrix $M$ for which the evaluation of the maximum $L(M)$ in Eq.~(\ref{LM}) is feasible on a standard desktop. Due to the numerical tests in Sec.~\ref{ssNum}, it is expected that $L(M)$ can be computed in a reasonable time up to a matrix dimension $n\approx90$ by running the B\&B algorithm.

In order to get $\vec a_i$, $i=1,\ldots,n$, we fix icosahedral symmetry of the set of vectors $\vec a_i$ and form a set of $2n$ vectors $\vec A_{2i-1}=\vec a_i$, $\vec A_{2i}=-\vec a_i$, for $i=1,\ldots,n$. Then we optimize the $2n$ unit vectors $\vec A_i$ in the $d$-dimensional Euclidean space assuming icosahedral symmetry such that the optimized configuration corresponds to a (local) minimum of the energy term
\begin{equation}
\label{energy}
E=\sum_{1\le i<j\le 2n}\frac{1}{\|\vec A_i-\vec A_j\|}.
\end{equation}
The goal of this optimization is to find an arrangement of vectors
$\vec A_i$ on the $(d-1)$-sphere, which distribute the sphere on
a relatively even manner. Minimization has been performed using a heuristic
search, the so-called Amoeba method~\cite{NM}. Given the fixed
arrangement of vectors $\vec a_i$ on Alice's side coming from the
above numerical search, due to symmetry reasons we pick the same
vectors on Bob's side. That is, we choose $\vec b_i=\vec a_i$ for
all $i=1,\ldots,n$. This gives an $n\times n$ correlation matrix
$C$ defined by the entries $C_{i,j}=\vec a_i\cdot\vec b_j$. Note
that all diagonal entries of this matrix are 1. In the next
subsections, we present two different methods to obtain the matrix
$M$ given the matrix $C$, considerably improving the lower bound
values of $K_G(d)$ for $d=3$ and $d=4$.

\subsection{$K_G(4)\ge 1.4731$ using a trial and error method}
\label{ssKG4dir}

We fix $n=92$ and generate $\vec a_i=\vec b_i$ in $d=4$ by
optimizing the energy~(\ref{energy}), from which we get the matrix
$C$ with entries
\begin{equation}
\label{EQ}
C_{i,j}=\vec a_i\cdot\vec b_j.
\end{equation}
With this in hand, we have to choose the form of the $n\times n$
matrix $M$. First we would like to demand that the matrix entries
$M_{i,j}$ are some function of the entries $C_{i,j}$. Hence we
define them as
\begin{equation}
\label{Mij}
M_{i,j}=[f(C_{i,j})],
\end{equation}
where we choose the form of the periodic function $f$ as
\begin{equation}
\label{fq}
f(q)=80\sin(\pi q/2)+100\sin(3\pi q/2),
\end{equation}
and $[x]$ denotes the nearest integer to $x$. Rounding has been
introduced in order for the entries of $M$ to be integer. On the
other hand, the constants appearing in the function~(\ref{fq}) are
chosen by trial and error such that they would provide good
performance, i.e., large lower bound values for $K_G(4)$. The next
subsection~\ref{ssKG4dist}, which uses the distance method to
lowerbound $K_G(4)$, will also shed light on the specific choice of
the function~(\ref{fq}).

Using the function in Eq.~(\ref{fq}), explicit calculations give
$Q(M,4)=\sum_{i,j}{M_{i,j}C_{i,j}}\simeq2.6785\times 10^5$ in
Eq.~(\ref{QMd}). On the other hand, $L(M)=181818$ coming from our
B\&B algorithm, which took roughly three months to evaluate it on
a desktop computer. Then the lower bound of $K_G(4)\ge
Q(M,4)/L(M)\simeq 1.4731$ follows from formula~(\ref{KGlower}). A
Mathematica file provides all the details of the matrices involved
in the computation~\cite{github}. It is noted that the algorithm
has very low memory requirements. We also recall that the $L(M)$
value does not depend on a specific ordering of the rows and
columns of $M$. In this respect, we found that the running time is
quite sensitive to the ordering of the rows, and it is worth
trying different orderings to improve time efficiency. We next
present an improved lower bound to $K_G(4)$, which uses the
distance method~\cite{gilbert,modgilbert} to generate the function
$f$ in Eq.~(\ref{Mij}).

\subsection{$K_G(4)\ge 1.4821$ using the distance method}
\label{ssKG4dist}

Here we give a specific $M$ matrix using the Gilbert's distance
method~\cite{gilbert,modgilbert}. In this way we get further
improvement on the lower bound to $K_G(4)$ presented in the
previous subsection.

Let us first briefly describe Gilbert's distance algorithm. It estimates
the distance between a point $P$ and an arbitrary convex set $S$
in some finite-dimensional Euclidean space via calls to an oracle
which performs linear optimizations over $S$. In our particular
case, the point is given by $P=v C$, that is, the correlation
matrix $C$ in Eq.~(\ref{EQ}) multiplied by a factor $0<v\le 1$.
The convex set in our case is the so-called $\pm 1$-polytope,
which is defined by the convex hull of its vertices as follows.
For a given $n$, the dimension of the polytope is $n\times n$, and
vertices $D_{\lambda}$ are given by matrices with entries
$D_{\lambda}(i,j)=a_ib_j$, where $\lambda$ corresponds to a
specific assignment of $a_i=\pm1$, $i=1,\ldots,n$ and $b_j=\pm1$,
$j=1,\ldots,n$. This amounts to $2^{2n-1}$ distinct vertices
$D_{\lambda}$. Any point inside the polytope is a convex
combination of vertices $D_{\lambda}$ with positive weights
$p(\lambda)$.

The factor $v$ is chosen in such a way that point $P=v C$ lies
(slightly) outside the $\pm 1$-polytope. To this end, let us
choose $C$ from Eq.~(\ref{EQ}) along with $v^*
=1/1.4731\simeq0.6788$, where $1.4731$ corresponds to the lower
bound $K_G(4)\ge 1.4731$ obtained in the preceding subsection. By
definition the point $v^*C$ is outside the $\pm 1$ polytope. Then
we call the distance algorithm~\cite{gilbert,modgilbert} where the
inputs to the problem are the point $v^*C$ and the description of
the $\pm 1$-polytope. The algorithm outputs (an estimate to) the
distance between the point $v^* C$ and the $\pm 1$-polytope by
providing a separating hyperplane with norm $M$, which is
identified with the $n\times n$ matrix $M$ that we are looking for.

The obtained matrix $M$ (after rounding to integers) is given in a
Mathematica file in the webpage~\cite{github}. Explicit
calculations show that
$Q(M,4)=\sum_{i,j}{M_{i,j}C_{i,j}}\simeq6.2223\times 10^8$ in
Eq.~(\ref{QMd}). On the other hand, the B\&B algorithm evaluates
$L(M)=419810256$, which took about three months on our desktop computer. Put
together, we get from formula~(\ref{KGlower}) the improved lower
bound $K_G(4)\ge Q(M,4)/L(M)\simeq 1.4821$.

\begin{figure}[tb!]
%\centering
%\includegraphics[width=1.3\columnwidth]{fig1.png}
% [trim={<left> <lower> <right> <upper>}]
\includegraphics[width=1.0\columnwidth,trim={2.5cm 8.0cm 2.5cm 8.0cm},clip]{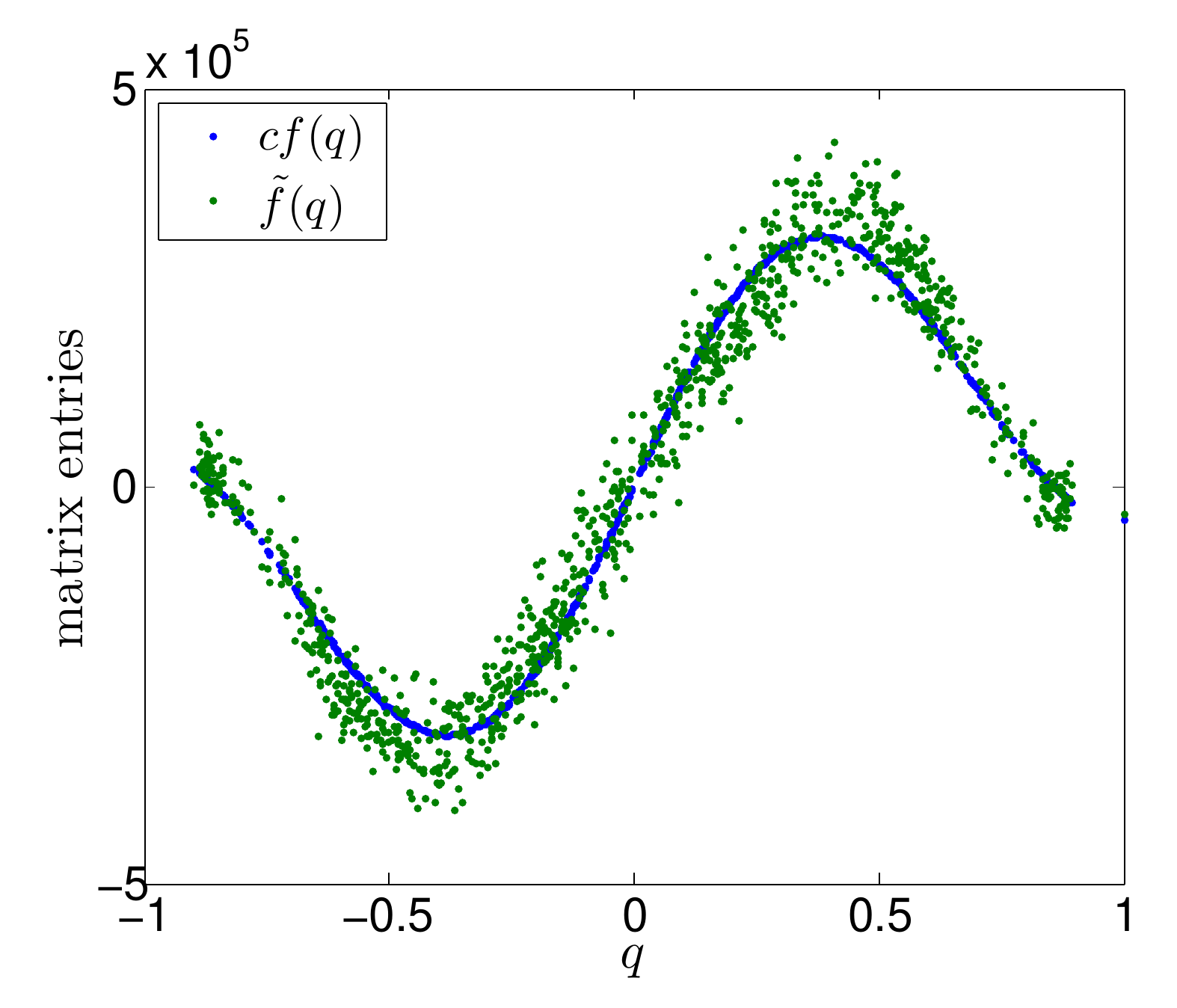}
\caption{The functions $f(q)$ and $\tilde f(q)$ are displayed in blue and green dots, respectively. An (irrelevant) multiplicative constant $c=2200$ is introduced for better comparison of the two plots. The feasible $q$ values on the $x$ axis correspond to the relation $q=\vec a_i\cdot \vec b_j$. }
\label{fig2}
\end{figure}

In the actual implementation of the distance algorithm, we
projected the problem from the space of $n\times n$ matrices to a
smaller subspace such that the entries $M_{i,j}$ of $M$ are given by $M_{i,j}=\tilde f(C_{i,j})$. In this way,
one can compare the two functions $f$ and $\tilde f$, where $f$ is
given by equation~(\ref{fq}) and $\tilde f$ results from the
distance algorithm in the present subsection. The two functions
are shown in Fig.~\ref{fig2}. According to the figure, the blue
dots (representing $f$) readily well approximate the scattered
green dots (representing $\tilde f$) within the full range of $q$
and can be considered as a coarse-grained version of it. Using
$\tilde f$ compared to $f$ in the definition of matrix $M$ gives
us the improved lower bound $K_G(4)\ge 1.4821$ compared to
$K_G(4)\ge 1.4731$.

\subsection{$K_G(3)\ge 1.4359$ using the distance method}
\label{ssKG3dist}

We optimized the energy formula~(\ref{energy}) by running the
Amoeba method for $d=3$ and $n=90$, and fixing icosahedral
symmetry. In this way, we obtained the unit vectors $\vec a_i$,
$i=1,\ldots,n$, on the sphere. Then, similarly to
Sec.~\ref{ssKG4dist}, the distance algorithm was consulted to
compute matrix $M$. This matrix $M$, whose entries are rounded to
the closest integers, is given in a Mathematica
file~\cite{github}. With this $M$ and $C$, we have
$Q(M,3)=\sum_{i,j}{M_{i,j}C_{i,j}}\simeq4.6560\times 10^8$. On the
other hand, $L(M)=324230014$ due to the B\&B algorithm, where the
running time was two weeks on our desktop computer. Then
we get from~(\ref{KGlower}) the lower bound $K_G(3)\ge
Q(M,3)/L(M)\simeq 1.4359$. It is noted that this value gives the
best upper bound of $v=1/1.4359\simeq 0.6964$ on the critical
visibility of the two-qubit Werner states improving on recent upper bounds~\cite{werner,open19}.

\section{Link to Bell nonlocality}
\label{sBell}

The Grothendieck constant has a direct link to quantum nonlocality problems~\cite{bell64,bellreview,bellreview2}, which we discuss briefly below. A detailed survey of this connection can be found in Ref.~\cite{carlos}.

In a quantum Bell-like experiment two parties perform local measurements on a shared entangled state~\cite{bell64}. Let Alice and Bob share a state $\rho$ in $\mathbb{C}^D\otimes\mathbb{C}^D$ and perform two-outcome projective measurements described by observables $A_x$ and $B_y$, which are $D$-dimensional Hermitian matrices with eigenvalues $\{\pm 1\}$. Here, $x,y=1,\ldots,n$ label the measurement settings. Then the correlator, which is the expectation value of the product of Alice and Bob's $\pm 1$ outcomes, is
\begin{equation}
\label{cxy}
c_{x,y}=\tr{\left(\rho A_x\otimes B_y\right)}
\end{equation}
for given settings $x$ and $y$. Such correlations associated with XOR nonlocal games are frequently studied in the computer science literature~\cite{xor}.

Given a dimension $D$, one wonders if a set of correlators
$\{c_{x,y},\, x,y=1,\ldots,n\}$ in Eq.~(\ref{cxy}) is quantum
realizable with a state $\rho\in\mathbb{C}^D\otimes\mathbb{C}^D$
using arbitrary POVM measurements, and also allowing Alice and Bob
to share an arbitrary large amount of randomness. If it happens
not to be the case, we say that the set of correlators
$\{c_{x,y}\}$ is not $D$-dimensional quantum realizable. A
convenient tool to address this problem is the use of dimension
witnesses~\cite{dimwit}. In what follows, we show that our main
result $K_G(3)<K_G(4)$ implies a set of correlators $\{c_{x,y}\}$,
which are not two-dimensional quantum realizable. This result is
based on earlier works~\cite{Tsirelson87,acin06,vp08}, and the
argument is as follows.

Let us consider a matrix $M'$ and unit vectors $\vec
a'_x\in\mathbb{R}^4$ and $\vec b'_y\in\mathbb{R}^4$ in
formula~(\ref{grot1}) which give rise to the exact value of
$K_G(4)$. It is noted that though the exact value of $K_G(4)$ is
unknown, there must exist some matrix $M'$ (of possibly
infinite dimension $n$) and unit vectors $\vec a'_x$, $\vec b'_y$
in the four-dimensional Euclidean space which give rise to
$K_G(4)$.

Tsirelson~\cite{Tsirelson87} has shown that all correlators $c_{x,y}$ equal to dot products $\vec a_x\cdot\vec b_y$ of the unit vectors $\vec a_x, \vec b_y\in\mathbb{R}^4$, are realizable as observables,
\begin{align}
\label{AxBy}
A_x&=\sum_{i=1}^{4}a_{x,i}\gamma_i,\nonumber\\
B_y&=\sum_{i=1}^{4}b_{y,i}\gamma_i^t,
\end{align}
on a pair of maximally entangled four-dimensional quantum systems,
$\ket{\psi_4}=(1/2)\sum_{i=1}^4\ket{i}\ket{i}$. Here, $a_{x,i}$, $b_{y,i}$ are entries of the four-dimensional
unit vectors $\vec a_x, \vec b_y\in\mathbb{R}^4$, respectively, $t$ denotes the transposition, and  $\gamma_i$ above are chosen as follows:
\begin{align}
\gamma_1&=\sigma_x\otimes\openone,\nonumber\\
\gamma_2&=\sigma_y\otimes\openone,\nonumber\\
\gamma_3&=\sigma_z\otimes\sigma_x,\nonumber\\
\gamma_4&=\sigma_z\otimes\sigma_z.
\end{align}
The above $\gamma_i$ matrices are traceless, anticommuting and square to the identity. Due to these properties, $A_x, B_y$ are valid traceless observables: $\tr(A_x)=\tr(B_y)=0$ and $A_x^2=B_y^2=\openone$. On the other hand, one has
\begin{equation}
\label{cpxy}
c_{x,y}=\tr{\left(\ket{\psi_4}\bra{\psi_4}A_x\otimes B_y\right)}=\tr(A_x B_y^t)/4.
\end{equation}
Applying Eq.~(\ref{AxBy}), and noting that $\tr(\gamma_i\gamma_j^t)=4\delta_{i,j}$, where $\delta_{i,j}$ denotes the Kronecker delta, we further have
\begin{equation}
\label{cpxy2}
c_{x,y}=\tr(A_x B_y^T)/4=\sum_{i}a_{x,i}b_{y,i}=\vec a_x\cdot\vec b_y.
\end{equation}
Replacing $\vec a_x$ and $\vec b_y$ with the particular vectors $\vec a_x',\vec b_y'$, which leads to the exact value of $K_G(4)$, we obtain that the correlators $c_{x,y}'=\vec a_x'\cdot\vec b_y'$ are realizable as observables $A_x'$, $B_y'$ on the state $\ket{\psi_4}=(1/2)\sum_{i=1}^4\ket{i}\ket{i}$. We next show that this set of correlators $c_{x,y}'$ has no two-dimensional quantum representation, i.e., it cannot be realized using qubit systems. The proof exploits the
strict relation $K_G(3)<K_G(4)$, which we have proven in the
preceding sections. To this end, let us consider the linear
function $I$ on the correlators $c_{x,y}$ in equation~(\ref{cxy})
written as
\begin{equation}
\label{Icxy}
I=\sum M'_{x,y}c_{x,y},
\end{equation}
where $M'_{x,y}$ is defined by the matrix $M'$ which attains the exact value of $K_G(4)$ in (\ref{grot1}). Let us then denote by $I^{(2)}$ the maximum of $I$ it can take if the correlators $c_{x,y}$ come from two-dimensional quantum systems. It appears that $I^{(2)}$ is defined by
\begin{equation}
I^{(2)}=\max\sum_{x,y=1}^m M'_{x,y}\vec a_x\cdot\vec b_y,
\end{equation}
where maximization is over all three dimensional unit vectors $\vec a_x$ and $\vec b_y$ \cite{dimwit,vp08}. However, by definition~(\ref{grot1}), $I^{(2)}$ is upper bounded by $K_G(3)L(M')$, where the function $L$ is defined by equation~(\ref{LM}). Therefore, we have the chain of inequalities
\begin{equation}
\frac{I^{(2)}}{L(M')}\le K_G(3)<K_G(4)=\frac{\sum_{x,y}M'_{x,y}c'_{x,y}}{L(M')},
\end{equation}
where comparing the leftmost and the rightmost terms gives us the strict relation
\begin{equation}
I^{(2)}<\sum_{x,y}M'_{x,y}c'_{x,y}.
\end{equation}
This tells us that the expression~(\ref{Icxy}) cannot be saturated
by correlations originating from qubit systems. Hence, the above
example shows the existence of a witness, which detects dimension greater than two in the
particular case where the witness matrix $M'$ is associated with
the $K_G(4)$ value in equation~(\ref{grot1}). A similar argument
in Ref.~\cite{dimwit} has shown the existence of a qutrit witness
from the strict relation $K_G(3)<K_G$. Here we showed that it
suffices to consider a pair of four-dimensional quantum systems to
certify correlations beyond qubit. Note also that dimension
witnesses including  any finite dimension $D$ appeared in the
literature based on different methods; see
e.g.~\cite{dimwit1,dimwit2,dimwit3,dimwit4,dimwit5}. More recent
works~\cite{dimwitt1,dimwitt2,dimwitt3,dimwitt4} revealed further
intriguing properties of the restricted dimensional quantum sets.

\section{Discussion}
\label{sDisc}

We proved that $K_G(4)$ is strictly larger than $K_G(3)$. To this
end, we used the so-called branch-and-bound algorithm commonly
used in the solution of NP-hard problems. This allowed us to solve
the problem~(\ref{LM}) up to matrix sizes 92 on a standard desktop
PC. Further, due to the principles of the branch-and-bound
algorithm (i.e. the calculation of the bounds and the branching in
each node is independent), it is a natural idea to adapt the
algorithm to GPU, Grid computing, or FPGA.

As we have shown, our result is relevant in quantum nonlocality,
as one can construct a dimension witness for detecting dimension greater than two based
on the relation $K_G(3)<K_G(4)$. To the best of our knowledge,
this is the first application of the branch-and-bound technique in
the context of quantum nonlocality, particularly in XOR nonlocal
games. However, we expect that the presented algorithm in
combination with other powerful methods (such as Gilbert's
distance algorithm) may find applications beyond XOR nonlocal
games as well. Such possible tasks concern Bell nonlocality with
more inputs~\cite{moreinp1,moreinp2}, more outcomes~\cite{moreout}
or genuine nonlocality in the case of multipartite
settings~\cite{svet1,svet2}. Note a recent method~\cite{baccari}
based on the Navascues-Pironio-Acin hierarchy~\cite{npa}, which
tackles these problems in a different way. Combining the two
approaches may also lead to improvement in our bipartite setting.

It would also be interesting to adapt the branch-and-bound
technique to bound the so-called unsteerability limit in EPR
steering inequalities~\cite{danipaul1,danipaul2} with large number
of inputs. Finally, the algorithm is likely applicable in random
access codes~\cite{qras} or non-contextuality inequalities as
well~\cite{noncontext1,noncontext2} with large number of
input-output alphabets.

\emph{Acknowledgements.} We thank N. Brunner, N. Gisin, J.
Kaniewski, Y.C. Liang, and M. Navascu\'es for useful discussions.
Technical assistance from B. K\H{o}m\H{u}ves is gratefully
acknowledged. We acknowledge financial support from the Hungarian
National Research Fund OTKA (Grant No.~K111734).

%%%%%%%%%%%%%%%%%%%%%%%%%%%%%%%%   End Main Text
%%%%%%%%%%%%%%%%%%%%%%%%%%%%%%%%   Begin Bibliography

%\bibliographystyle{linksen}  %Bibliography style for texts in english (Author: Mateus Araujo maltusan@gmail.com)
%\bibliography{mtqbib}        %MTQ personal bibliography database

%%%%%%%%%%%%%%%%%%%%%%%%%%%%%%%%   Put here the .bib generated by bibtex in case only one .tex file is required
%\providecommand{\href}[2]{#2}\begingroup\raggedright\begin{thebibliography}{10}

%\pagebreak

%\section{Appendix A}
%\label{appA}

%\appendix
%\onecolumngrid
%\section{The biased CHSH inequality}
%\label{app:biased-chsh} In this Appendix we analyse

\appendix
\onecolumngrid
\section{Description and implementation details of the $K_{m,n}$ programming algorithm}
\label{appa}

\section*{Introduction}\label{introduction}

\(K_{m,n}\)-quadratic programming is a quadratic optimization
problem with binary variables. In the main text the algorithm to
solve \(K_{m,n}\)-quadratic programming is defined (where we have set $m=n$). In this appendix,
we prove the correctness of the algorithm. We also provide tips about the efficient
implementation of the algorithm on a desktop computer. An implementation of this algorithm with application in XOR nonlocal games is publicly available at~\cite{github}.
%{[}\url{https://github.com/divipp/kmn-programming}{]}.

\section*{Notation}\label{notation}

\begin{longtable}[]{@{}ll@{}}
\toprule \(\mathbb{N}\) & set of natural numbers\tabularnewline
\(\mathbb{Z}\) & set of integers\tabularnewline \(A^n\) &
\(n\)-ary Cartesian product \(A\times\cdots\times
A\)\tabularnewline \(v_i\) & \(i\)th coordinate of \(v\in A^n,
i=1,2,\dots,n\)\tabularnewline \((v_1,v_2,\dots,v_n)\) &
construction of \(v\in A^n\)\tabularnewline \(\|v\|_1\) &
Manhattan norm, i.e. \(\sum_i |v_i|\)\tabularnewline
\(\mat{n}{m}(A)\) & matrices with \(n\) rows and \(m\) columns
over \(A\)\tabularnewline \(M_i\) & \(i\)th row of \(M\in
\mat{n}{m},\; i = 1,2,\dots,n\)\tabularnewline \bottomrule
\end{longtable}

\section*{\texorpdfstring{\(K_{m,n}\)-quadratic
programming}{K\_\{m,n\}-quadratic
programming}}\label{k_mn-quadratic-programming}

Let \(M\) be an \(n\times m\) matrix of integers. The goal of
\(K_{m,n}\)-quadratic programming is to efficiently compute the
\(\opt\) function, which is defined as follows:

\begin{longtable}[]{@{}l@{}}
\toprule \(\opt:\mat{n}{m}(\Z)\rightarrow \Z\)\tabularnewline
\(\displaystyle\opt(M) \coloneqq \max_{a_i=\pm 1, b_j=\pm 1}
\sum_{i=1}^n\sum_{j=1}^m M_{ij}a_i b_j\)\tabularnewline
\bottomrule
\end{longtable}

\subsection*{Basic properties}\label{basic-properties}

\emph{Theorem}
\[\opt(M) = \max_{a_i=\pm 1} \left\|\sum_{i=1}^n a_i M_i\right\|_1.\]

\emph{Proof} \[\begin{array}{ll} \max_{a_i=\pm 1}
\left\|\sum_{i=1}^n a_i M_i\right\|_1
&= \max_{a_i=\pm 1} \sum_{j=1}^m \left|\sum_{i=1}^n a_i M_{ij}\right| \\
&= \max_{a_i=\pm 1}\max_{b_j=\pm 1} \sum_{j=1}^m b_j\left(\sum_{i=1}^n a_i M_{ij}\right) \\
&= \max_{a_i=\pm 1, b_j=\pm 1} \sum_{i=1}^n\sum_{j=1}^m M_{ij}a_i b_j \\
&= \opt(M).
\end{array}\]

\emph{Theorem}

Let \(M\in\mat{n}{m}, M=(M_1,M_2,\dots,M_n)\), where \(M_i\) is the
\(i\)th row of \(M\). \[\opt(M) = \max(\opt(M^{+}),
\opt(M^{-})),\] where \(M^{+}=(M_1+M_2,M_3,M_4,\dots,M_n)\) and
\(M^{-}=(M_1-M_2,M_3,M_4,\dots,M_n).\)

\emph{Proof} \[\begin{array}{ll} \opt(M)
&= \max_{a_i=\pm 1} \left\|\sum_{i=1}^n a_i M_i\right\|_1 \\
&= \max(\max_{a_i=\pm 1, a_1=a_2} \left\|\sum_{i=1}^n a_i M_i\right\|_1, \max_{a_i=\pm 1, a_1\ne a_2} \left\|\sum_{i=1}^n a_i M_i\right\|_1) \\
&= \max(\opt(M^{+}), \opt(M^{-})).
\end{array}\]

\emph{Theorem}

Let \(M\in\mat{n}{m}, M=(M_1,M_2,\dots,M_n)\), where \(M_i\) is the
\(i\)th row of \(M\). \[\opt(M) \le \opt(M^{U}) + \opt(M^{L}).\]
where \(M^{U}=(M_1,M_2,\dots,M_k)\) and
\(M^{L}=(M_{k+1},M_{k+2},\dots,M_n).\)

\emph{Proof} \[\begin{array}{ll} \opt(M)
&= \max_{a_i=\pm 1} \left\|\sum_{i=1}^n a_i M_i\right\|_1 \\
&\le \max_{a_i=\pm 1} \left(\left\|\sum_{i=1}^k a_i M_i\right\|_1 + \left\|\sum_{i=k+1}^{n} a_i M_i\right\|_1\right) \\
&= \max_{a_i=\pm 1} \left\|\sum_{i=1}^k a_i M_i\right\|_1 + \max_{a_i=\pm 1} \left\|\sum_{i=k+1}^{n} a_i M_i\right\|_1 \\
&= \opt(M^{U}) + \opt(M^{L}).
\end{array}\]

\section*{\texorpdfstring{Recursive calculation of
\(\opt\)}{Recursive calculation of
\textbackslash{}opt}}\label{recursive-calculation-of-opt}

We define a recursive function \(f\) to calculate \(\opt\). The
function \(f\) is not efficient, but it helps to understand the
efficient functions defined later and it is also used in their
correctness proofs.

Let \(n, m \in \mathbb{N}\).\\
Let \(M = (M_1,M_2,\dots,M_n) \in \mat{n}{m}\).

\(M\) is a fixed parameter of the function \(f\) so it is placed
in the subscript as \(f_M\).

The recursive function \(f_M\) is defined as

\begin{longtable}[]{@{}l@{}}
\toprule \(f_M:\mathbb{N}\times \Z^m
\rightarrow\Z\)\tabularnewline \(f_M(k,v) \coloneqq
\left\{\begin{array}{ll} \|v\|_1 & \text{if } k = n, \\
\max(f_M(k+1, v+M_{k+1}), f_M(k+1, v-M_{k+1})) & \text{otherwise}.
\end{array}\right.\)\tabularnewline \bottomrule
\end{longtable}

\emph{Theorem} \[f_M(k,v) = \opt((v,M_{k+1},M_{k+2},\dots,M_n)).\]

\emph{Proof}

By induction on \(k = n, n-1, n-2, \dots\):

\begin{itemize}
\tightlist \item
  Base case: \(k = n\)
  \[f_M(k,v) = \|v\|_1 = \opt((v)) = \opt((v,M_{k+1},\dots,M_n)).\]
\item
  Inductive step: \(k < n\) \[\begin{array}{ll}
  f_M(k,v)
  &= \max(f_M(k+1, v+M_{k+1}), f_M(k+1, v-M_{k+1})) \\
  &= \max(\opt((v+M_{k+1},M_{k+2},\dots,M_n)), \opt((v-M_{k+1},M_{k+2},\dots,M_n))) \\
  &= \opt((v,M_{k+1},M_{k+2},\dots,M_n)).
  \end{array}\]
\end{itemize}

\emph{Corollary} \[\opt(M) = f_M(1, M_1).\]

\emph{Example}

Let \(M = \left(\begin{array}{rrrr}2& 3& 3& 0\\ 3& 2& -3& -3\\ 3&
-3& 2& 3\\ 0& -3& 3& 2\end{array}\right) \in \mat{4}{4}\).
\[\begin{array}{ll}
\opt(M)&= f_M(1,(2,3,3,0)) \\
&= \max(f_M(2,(5,5,0,-3)),f_M(2,(-1,1,6,3))) \\
&= \max(\max(f_M(3,(8,2,2,0)),f_M(3,(2,8,-2,-6))) \\
&\phantom{= \max},\max(f_M(3,(2,-2,8,6)),f_M(3,(-4,4,4,0)))) \\
&= \max(\max(\max(f_M(4,(8,-1,5,2)),f_M(4,(8,5,-1,-2))) \\
&\phantom{= \max(\max},\max(f_M(4,(2,5,1,-4)),f_M(4,(2,11,-5,-8)))) \\
&\phantom{= \max}, \max(\max(f_M(4,(2,-5,11,8)),f_M(4,(2,1,5,4))) \\
&\phantom{= \max(\max},\max(f_M(4,(-4,1,7,2)),f_M(4,(-4,7,1,-2))))) \\
&= \max(\max(\max(\|(8,-1,5,2)\|_1,\|(8,5,-1,-2)\|_1) \\
&\phantom{= \max(\max},\max(\|(2,5,1,-4)\|_1,\|(2,11,-5,-8)\|_1)) \\
&\phantom{= \max}, \max(\max(\|(2,-5,11,8)\|_1,\|(2,1,5,4)\|_1) \\
&\phantom{= \max(\max},\max(\|(-4,1,7,2)\|_1,\|(-4,7,1,-2)\|_1))) \\
&= \max(\max(\max(16,16) \\
&\phantom{= \max(\max},\max(12,26)) \\
&\phantom{= \max},\max(\max(26,12) \\
&\phantom{= \max(\max},\max(14,14))) \\
&= 26.
\end{array}\]

Note that the total number of \(f_M(k,v)\) calls is \(1 + 2 + 4 +
\cdots + 2^{n-1} = 2^n-1 = 15\).

\section*{Speeding up the recursion}\label{speeding-up-the-recursion}

We define another recursive function, \(g\), to calculate
\(\opt\). \(g\) is more efficient than \(f\) because it tries to
skip whole branches of recursive calls by comparing the best found
maximum so far and the estimated result of the branch.

Let \(c_k \ge \opt((M_{k+1},M_{k+2},\dots,M_n))\) arbitrary
constants, where \(k = 1, 2, \dots, n-1\). We discuss later how to
choose \(c_k\).
Let \(c_n = 0\).\\
Let \(c=(c_1, c_2, \dots, c_n)\in\Z^n\).

\(M\) and \(c\) are fixed parameters of the function \(g\) so they
are placed in the subscript as \(g_{M,c}\).

\(g_{M,c}\) is defined as

\begin{longtable}[]{@{}l@{}}
\toprule \(g_{M,c}:\mathbb{N}\times \Z^m\times \Z
\rightarrow\Z\)\tabularnewline \(g_{M,c}(k,v,m) \coloneqq
\left\{\begin{array}{ll} m & \text{if } m \ge \|v\|_1 + c_k, \\
\|v\|_1 & \text{otherwise if } k = n, \\ g_{M,c}(k+1, v-M_{k+1},
g_{M,c}(k+1, v+M_{k+1}, m)) & \text{otherwise}.
\end{array}\right.\)\tabularnewline \bottomrule
\end{longtable}

\emph{Theorem} \[g_{M,c}(k,v,m) = \max(f_M(k,v), m).\]

\emph{Proof}

By induction on \(k = n, n-1, n-2, \dots\):

\begin{itemize}
\tightlist \item
  Base case: \(k = n\)

  \begin{itemize}
  \tightlist
  \item
    Case \(m \ge \|v\|_1\)
    \[g_{M,c}(k,v,m) = m = \max(\|v\|_1, m) = \max(f_M(k,v), m).\]
  \item
    Case \(m < \|v\|_1\)
    \[g_{M,c}(k,v,m) = \|v\|_1 = \max(\|v\|_1, m) = \max(f_M(k,v), m).\]
  \end{itemize}
\item
  Inductive step: \(k < n\)

  \begin{itemize}
  \tightlist
  \item
    Case \(m \ge \|v\|_1 + c_k\) \[\begin{array}{ll}
    g_{M,c}(k,v,m)
    &= m \\
    &= \max(\|v\|_1 + c_k, m) \\
    &= \max(\opt((v,M_{k+1},\dots,M_n)), m) \\
    &= \max(f_M(k,v), m).
    \end{array}\]
  \item
    Case \(m < \|v\|_1 + c_k\) \[\begin{array}{ll}
    g_{M,c}(k,v,m)
    &= g_{M,c}(k+1, v-M_{k+1}, g_{M,c}(k+1, v+M_{k+1}, m)) \\
    &= \max(f_M(k+1, v-M_{k+1}), \max(f_M(k+1, v+M_{k+1}), m)) \\
    &= \max(\max(f_M(k+1, v-M_{k+1}), f_M(k+1, v+M_{k+1})), m) \\
    &= \max(f_M(k, v), m).
    \end{array}\]
  \end{itemize}
\end{itemize}

\emph{Corollary} \[\opt(M) = g_{M,c}(1,M_1, 0).\]

\subsection*{\texorpdfstring{Choosing
\(c_k\)}{Choosing c\_k}}\label{choosing-c_k}

The \(c_k\) constants can be chosen arbitrarily unless they are
greater than or equal to \(\opt((M_{k+1},M_{k+2},\dots,M_n))\). Lower
\(c_k\) constants prevent more \(g_{M,c}(k,v,m)\) computations.
There is a trade-off between computing the lower bound of \(c_k\)
to speed up later computations or using less resources on \(c_k\)
and doing more computation later.

The lower bound of \(c_k\) can alo be computed with \(g_{M,c}\):
\[\opt((M_{k+1},M_{k+2},\dots,M_n)) = g_{M,c}(k+1,M_{k+1}, 0).\]

We have found by experience that it is worthwhile to compute the lower
bound of \(c_n, c_{n-1}, \dots, c_i\), and set the remaining
\(c_{i-1}, c_{i-2}, \dots, c_2\) constants to \(\infty\), where
\(i\) is around \(\lceil n/4 \rceil\). Note that during the
computation of the lower bound of \(c_k\), the \(c_j, j>k\)
constants are also needed, so one should compute the lower bounds
of \(c_n, c_{n-1}, \dots, c_i\) one after another in this order.

\emph{Example}

Let \(M = \left(\begin{array}{rrrr}2& 3& 3& 0\\ 3& 2& -3& -3\\ 3&
-3& 2& 3\\ 0& -3& 3& 2\end{array}\right) \in \mat{4}{4}\).

\[\begin{array}{ll}
c_4 &= 0 \comm{by definition} \medskip\\
c_3 &\coloneqq \opt((M_4)) \comm{choose the lower bound} \\
       &= g_{M,c}(4,(0,-3,3,2),0) \\
       &= \|(0,-3,3,2)\|_1 \comm{because $0 \not\ge \|(0,-3,3,2)\|_1 + c_4$} \\
       &= 8 \medskip\\
c_2 &\coloneqq \infty \comm{avoid computation of $\opt((M_3, M_4))$} \medskip\\
c_1 &\coloneqq \infty \comm{avoid computation of $\opt((M_2, M_3, M_4))$} \medskip\\
\opt(a) &= g_{M,c}(1,(2,3,3,0),0) \\
        &= g_{M,c}(2,(-1,1,6,3),g_{M,c}(2,(5,5,0,-3),0)) \\
        &= g_{M,c}(2,(-1,1,6,3),g_{M,c}(3,(2,8,-2,-6),g_{M,c}(3,(8,2,2,0),0))) \\
        &= g_{M,c}(2,(-1,1,6,3),g_{M,c}(3,(2,8,-2,-6),g_{M,c}(4,(8,5,-1,-2),g_{M,c}(4,(8,-1,5,2),0)))) \\
        &= g_{M,c}(2,(-1,1,6,3),g_{M,c}(3,(2,8,-2,-6),g_{M,c}(4,(8,5,-1,-2),\|(8,-1,5,2)\|_1))) \\
        &= g_{M,c}(2,(-1,1,6,3),g_{M,c}(3,(2,8,-2,-6),g_{M,c}(4,(8,5,-1,-2),16))) \\
        &= g_{M,c}(2,(-1,1,6,3),g_{M,c}(3,(2,8,-2,-6),16)) \\
        &= g_{M,c}(2,(-1,1,6,3),g_{M,c}(4,(2,11,-5,-8),g_{M,c}(4,(2,5,1,-4),16))) \\
        &= g_{M,c}(2,(-1,1,6,3),g_{M,c}(4,(2,11,-5,-8),16)) \\
        &= g_{M,c}(2,(-1,1,6,3),\|(2,11,-5,-8)\|_1) \\
        &= g_{M,c}(2,(-1,1,6,3),26) \\
        &= g_{M,c}(3,(-4,4,4,0),g_{M,c}(3,(2,-2,8,6),26)) \\
        &= g_{M,c}(3,(-4,4,4,0),26) \comm{optimization kicks in}\\
        &= 26. \comm{optimization kicks in}
\end{array}\]

Note that the total number of \(g_{M,c}(k,v,m)\) calls is \(12\).

\section*{Tail-recursive form}\label{tail-recursive-form}

It is possible to refactor \(g_{M,c}\) into two mutually
tail-recursive functions \(d_{M,c}\) and \(u_{M,c}\) such that
each recursive call is a tail call, i.e.,~there are no further
operations involved after the call is completed~\cite{tail}. Tail
calls can be implemented by goto statements so they do not need
stack operations, which is a requirement on a GPU and also speeds
up computation on a CPU.

The definitions of \(d_{M,c}\) and \(u_{M,c}\) are

\begin{longtable}[]{@{}l@{}}
\toprule \(d_{M,c}, u_{M,c}: \mathbb{N}\times \mathbb{N}\times
\Z^m\times \mathbb{N} \rightarrow\mathbb{N}\)\tabularnewline
\(d_{M,c}(k,b,v,m) \coloneqq \left\{\begin{array}{ll}
u_{M,c}(k,b,v,m) & \text{if } m \ge \|v\|_1 + c_k, \\
u_{M,c}(k,b,v,\|v\|_1) & \text{otherwise if } k = n, \\
d_{M,c}(k+1, 2b, v+M_{k+1}, m) & \text{otherwise}.
\end{array}\right.\medskip\)\tabularnewline \(u_{M,c}(k,b,v,m)
\coloneqq \left\{\begin{array}{ll} m & \text{if } k = 1, \\
d_{M,c}(k,b+1,v-2M_k,m) & \text{otherwise if } b = 2b', \\
u_{M,c}(k-1,b',v+M_k,m) & \text{otherwise if } b = 2b'+1.
\end{array}\right.\)\tabularnewline \bottomrule
\end{longtable}

\emph{Theorem}

For all \(k\in\mathbb{N}, 1\le k\le n\) and \(a_2,a_3,\dots,a_k=
\pm 1\),

\[d_{M,c}\Big(k,\digits{a_2 a_3 \cdots a_k},M_1 + \sum_{i=2}^k a_i M_i, \max_{\scriptsize\begin{array}{c}b_2,b_3,\dots,b_n= \pm 1\\ \digits{b_2 b_3 \cdots b_k}<\digits{a_2 a_3 \cdots a_k}\end{array}} \big(M_1+\sum_{i=2}^n b_i M_i\big)\Big)
= \opt(M),\]
\[u_{M,c}\Big(k,\digits{a_2 a_3 \cdots a_k},M_1 + \sum_{i=2}^k a_i M_i, \max_{\scriptsize\begin{array}{c}b_2,b_3,\dots,b_n= \pm 1\\ \digits{b_2 b_3 \cdots b_k}\le\digits{a_2 a_3 \cdots a_k}\end{array}} \big(M_1+\sum_{i=2}^n b_i M_i\big)\Big)
= \opt(M),\] where
\[\digits{x_1 x_2 \cdots x_i} \coloneqq \sum_{j=1}^i \frac{1-x_j}{2}2^{i-j},\]
and \[\max_{x\in\emptyset} x = 0.\]

\emph{Sketch of the proof}\\
One has to show that if the parameters of \(d_{M,c}\) and
\(u_{M,c}\) are in the form given in the theorem, then in each
possible case, the next call of \(d_{M,c}\) or \(u_{M,c}\) has
parameters in the form given in the theorem too.

\emph{Corollary} \[\opt(M) = d_{M,c}(1, 0, M_1, 0).\]

\emph{Example}

Let \(M = \left(\begin{array}{rrrr}2& 3& 3& 0\\ 3& 2& -3& -3\\ 3&
-3& 2& 3\\ 0& -3& 3& 2\end{array}\right) \in \mat{4}{4}\).

Let \(c_4 = 0, c_3 = 8, c_2 = \infty, c_1 = \infty\) as in the
previous example.

\[\begin{array}{lll}
\opt(M)&= d_{M,c}(1, 0, (2,3,3,0), &0) \\
&= d_{M,c}(2, 0, (5,5,0,-3), &0) \\
&= d_{M,c}(3, 0, (8,2,2,0), &0) \\
&= d_{M,c}(4, 0, (8,-1,5,2), &0) \comm{$\|(8,-1,5,2)\|_1=16$} \\
&= u_{M,c}(4, 0, (8,-1,5,2), &16) \comm{$0=2\cdot 0$}\\
&= d_{M,c}(4, 1, (8,5,-1,-2), &16) \\
&= u_{M,c}(4, 1, (8,5,-1,-2), &16) \comm{$1=2\cdot 0+1$}\\
&= u_{M,c}(3, 0, (8,2,2,0), &16) \comm{$0=2\cdot 0$}\\
&= d_{M,c}(3, 1, (2,8,-2,-6), &16) \\
&= d_{M,c}(4, 2, (2,5,1,-4), &16) \\
&= u_{M,c}(4, 2, (2,5,1,-4), &16) \comm{$2=2\cdot 1$}\\
&= d_{M,c}(4, 3, (2,11,-5,-8), &16) \comm{$\|(2,11,-5,-8)\|_1=26$} \\
&= u_{M,c}(4, 3, (2,11,-5,-8), &26) \comm{$3=2\cdot 1+1$}\\
&= u_{M,c}(3, 1, (2,8,-2,-6), &26) \comm{$1=2\cdot 0+1$}\\
&= u_{M,c}(2, 0, (5,5,0,-3), &26) \comm{$0=2\cdot 0$}\\
&= d_{M,c}(2, 1, (-1,1,6,3), &26) \\
&= d_{M,c}(3, 2, (2,-2,8,6), &26) \comm{$26\ge\|(2,-2,8,6)\|_1+8$}\\
&= u_{M,c}(3, 2, (2,-2,8,6), &26) \comm{$2=2\cdot 1$}\\
&= d_{M,c}(3, 3, (-4,4,4,0), &26) \comm{$26\ge\|(-4,4,4,0)\|_1+8$}\\
&= u_{M,c}(3, 3, (-4,4,4,0), &26) \comm{$3=2\cdot 1+1$}\\
&= u_{M,c}(2, 1, (-1,1,6,3), &26) \comm{$1=2\cdot 0+1$}\\
&= u_{M,c}(1, 0, (2,3,3,0), &26) \comm{$k=1$}\\
&= 26.
\end{array}\]

\end{document}